\documentclass[manuscript]{aastex}

\shorttitle{Estimation of Twist and Magnetic Energy}
\shortauthors{S. K. Tiwari et al.}

\begin{document}

\title{Effect of Polarimetric Noise on the Estimation of Twist
and Magnetic Energy of Force-Free Fields}

\author{Sanjiv Kumar Tiwari,
        P.~Venkatakrishnan,
        Sanjay~Gosain and
        Jayant Joshi }
\affil{Udaipur Solar Observatory, Physical Research Laboratory,\\
 Dewali, Bari Road, Udaipur-313 001, India}
\email{stiwari@prl.res.in}
\email{pvk@prl.res.in}
\email{sgosain@prl.res.in}
\email{jayant@prl.res.in}

\begin{abstract}
The force-free parameter $\alpha$, also known as helicity parameter
or twist parameter, bears the same sign as the magnetic helicity
under some restrictive conditions.
The single global value of $\alpha$ for a whole active
region gives the degree of twist per unit axial length.
We investigate the effect of polarimetric noise
on the calculation of global $\alpha$ value and magnetic energy of an
analytical bipole.
The analytical bipole has been generated using the force-free field
approximation with a known
value of constant $\alpha$ and magnetic energy.
The magnetic parameters obtained from the analytical bipole are used to
generate Stokes profiles from the Unno-Rachkovsky solutions for
polarized radiative transfer equations.
Then we add random noise of the order of 10$^{-3}$ of the continuum
intensity (I$_{c}$) in these profiles to simulate the
real profiles obtained by modern spectropolarimeters like
Hinode (SOT/SP), SVM (USO), ASP, DLSP, POLIS, SOLIS etc.
These noisy profiles are then inverted using a Milne-Eddington
inversion code to retrieve the magnetic parameters.
Hundred realizations of this process of adding random noise and
polarimetric inversion is repeated to study the distribution of error
in global $\alpha$ and magnetic energy values.
The results show that : (1). the sign of $\alpha$ is not influenced
by polarimetric noise and very accurate values
of global twist can be calculated, and
(2). accurate estimation of magnetic energy with uncertainty as low as
0.5\% is possible under the force-free condition.
\end{abstract}


\section{Introduction}

Helical structures in the solar features like sunspot whirls were
reported long back by George E. Hale in 1925
\citep{hale25,hale27}.
He found that about 80\% of the sunspot whirls were
counterclockwise in the northern hemisphere and clockwise
in the southern hemisphere.
Later, in 1941 the result was confirmed by Richardson
\citep{rich41} by extending
the investigation over four solar cycles. This hemispheric
pattern was found to be independent of the solar cycle.
Since the 90's, the subject has been rejuvenated and this
hemispheric behaviour independent of sunspot cycle is claimed to be
observed for many of the solar features like active regions
\citep{seeh90,pcm95,long98,
abra96,bao98,hagi05},
filaments \citep{mart94,pevt03,bern05},
coronal loops \citep{rust96,pevt01},
interplanetary magnetic clouds (IMCs)
\citep{rust94},
coronal X-ray arcades
\citep{mart96} and network magnetic fields
\citep{pcl01,pevt07} etc.

Helicity is a physical quantity that measures the degree of
linkages and twistedness in the field lines
\citep{berg84}.
Magnetic helicity H$_{m}$ is given by
a volume integral over the scalar product of the magnetic
field \textbf{B} and its vector potential \textbf{A}
\citep{els56}.

\begin{equation}\label{}
         H_{m} = \int_V {\bf A}\cdot{\bf B}~dV
\end{equation}
with \textbf{B}= $\nabla \times$ \textbf{A}.

It is well known that the vector potential \textbf{A }is not unique,
thereby preventing the calculation of a unique value for the magnetic
 helicity from the equation (1). \cite{seeh90} pointed out that the
helicity of magnetic field can best be characterized by the force-free
parameter $\alpha$, also known as the helicity parameter or twist
parameter. The force-free condition
\citep{chandra61,parker79} is given as,
\begin{equation}\label{}
    \nabla \times \bf B = \alpha \bf B
\end{equation}
Alpha is a measure of degree of twist per unit axial length
(see Appendix-A for details of physical meaning of alpha).
This is a local parameter which can
vary across the field but is constant along the field lines.

Researchers have claimed to have determined the
sign of magnetic helicity
on the photosphere by calculating alpha, e.g.
$\alpha_{best}$ \citep{pcm95},
averaged alpha e.g.  $<\alpha_{z}>$ = $<{J_{z}} / {B_{z}}>$
\citep{pcm94} with current density
${J_{z}} = ({\bf{\nabla\times B}})_{z}$.
Some authors have used current helicity density
$H_{c }= {B_{z}} \cdot {J_{z}}$ and $\alpha_{av}$
\citep{bao98,hagi04,hagi05}.
A good correlation was found between  $\alpha_{best}$ and
$\langle\alpha_{z}\rangle$ by \cite{burn04} and
\cite{leka96}.
But the sign of magnetic helicity cannot be inferred from the
force-free parameter $\alpha$ under all
conditions (See Appendix B).

It is well known that the reliable measurements of vector
magnetic fields are needed to study various important
parameters like electric currents in the active regions,
magnetic energy dissipation during flares, field geometry
of sunspots, magnetic twist etc. The study of error
propagation from polarization measurements to vector field
parameters is very important \citep{lites85,klim92}.
\cite{klim92} have studied the effects of realistic
errors e.g., due to random polarization noise, crosstalk
between different polarization signals, systematic polarization
bias and seeing induced crosstalk etc. on known magnetic fields.
They derived analytical expressions
for how these errors produce errors in the estimation of magnetic
energy (calculated from virial theorem).
However, they simulated these effects for magnetographs
which sample polarization at few fixed wavelength positions in
line wings. It is well known that such observations lead to
systematic under-estimation of field strength and also suffer
from magneto-optical effects \citep{west83}.
Whereas in our analysis, we simulate the effect of polarimetric
 noise on field parameters as deduced by full Stokes inversion.
The details are discussed in the section 6.

\cite{pcm95} found large variations in the
global $\alpha$ values from repeated observations of the
 same active regions.
It is important to model the
measurement uncertainties before looking for physical
explanations for such a scatter.

In a study by \cite{hagy99a} the noise levels in
the observed fields were analyzed, but a
quantitative relationship between the uncertainties in fields
and the uncertainties in global $\alpha$ value were not
established. They could only determine the extent to which
the incremental introduction of noise affects the observed
value of alpha. However, for the proper tracking of error
propagation, we need to start with ideal data devoid
of noise and with known values of $\alpha$ and magnetic energy.
We follow the latter approach in our present analysis.

Here, we estimate the accuracy in the calculation of the $\alpha$
parameter and the magnetic energy due to
different noise levels in the spectro-polarimetric profiles.
Modern instruments measure the
full Stokes polarization parameters within the line profile.
Basically there are two types of spectro-polarimeters :
(i) Spectrograph based e.g.,
Advanced Stokes Polarimeter (ASP : \cite{elm92}),
Zurich Imaging Polarimeter (ZIMPOL : \cite{kel92,povel95,sten96,sten97}),
THEMIS-MTR (\cite{arnaud98}),
SOLIS - Vector Spectro-Magnetograph (VSM : \cite{jones02,kel03}),
Polarimetric Littrow Spectrograph (POLIS : \cite{schm03}),
Diffraction Limited Spectro-polarimeter (DLSP :
\cite{sankar04,sankar06}),
Hinode (SOT/SP : \cite{tsun08}), etc. and
(ii) Filter-based e.g.,
Imaging Vector Magnetograph (IVM) at
Mees Solar Observatory, Hawaii \citep{mick96},
Solar Vector Magnetograph at Udaipur Solar
Observatory (SVM-USO)\citep{gosain04,gosain06} etc.

Earlier magnetographs like Crimea \citep{step62},
MSFC \citep{hagy82},
HSP \citep{mick85}, OAO \citep{maki85},
HSOS \citep{ai86}, Potsdam vector magnetograph \citep{stau91},
SFT \citep{saku95} etc.
were mostly based on polarization measurements at a
few wavelength positions in the line wings and hence subject to
Zeeman saturation effects as well as magneto-optical effects like
Faraday rotation \citep{west83,hagy00}.

The magnetic field vector deduced from Stokes profiles by
modern techniques are almost free
from such effects \citep{skum87,sanch98,socas01}.

This paper serves three purposes. First, we estimate the
 error in the calculation of field strength, inclination
and azimuth and thereafter in the
calculation of the vector field components ${B}_{x}$, ${B}_{y}$
and ${B}_{z}$.
Second, we estimate the error in the determination of
global $\alpha$ due to noise in polarimetric profiles constructed
from the analytical vector field data.
Third, we estimate the error in the calculation of
magnetic energy derived using virial theorem,
due to polarimetric noise.

In the next section (section 2) we discuss a direct method for
calculation of a single global $\alpha$ for an active region.
In section 3, we describe the method of simulating an analytical
bipole field. Section 4 contains the analysis and the results.
Error estimation in global $\alpha$ is given in section 5.
In section 6 we discuss the process of estimating the error in
the virial magnetic energy.
Section 7 deals with discussion and conclusions.

\section{A direct method for calculation of global $\alpha$}

Taking the z-component of magnetic field, from the force-free field
equation (2) $\alpha$ can be written as,
\begin{equation}\label{}
    \alpha = \frac{(\nabla \times\bf B)_z} {B_z}
\end{equation}
For a least squares minimization, we should have
\begin{eqnarray}\label{}
 \nonumber   \sum(\alpha - \alpha_{g})^2 &=& minimum\\
  or,~~~~~~~~~~~~~      \alpha_{g}  &=& (1/N) \sum \alpha
\end{eqnarray}

where $\alpha$ is the local value at each pixel, $\alpha_{g}$
is the global value of $\alpha$ for the complete active region and N is
total number of pixels.\\
Since eqn.(4) will lead to singularities at the neutral lines
where B$_z$ approaches 0, therefore the next moment of minimization,
\begin{equation}\label{}
    \sum(\alpha - \alpha_{g})^2  B_z^2 = minimum
\end{equation}
should be used.
From eqn.(5) we have
\begin{equation}\label{}
\frac{\partial}{\partial\alpha_{g}}(\sum(\alpha - \alpha_{g})^2  B_z^2) = 0
\end{equation}
which leads to the following result,
\begin{equation}\label{}
\alpha_{g}=\frac{\sum(\frac{\partial B_y}{\partial x} -
\frac{\partial B_x}{\partial y})B_z}{\sum B_z^2}
\end{equation}
This formula gives a single global value of $\alpha$ in a
sunspot and is the same as $\alpha_{av}^{(2)}$ of \cite{hagi04}.
We prefer this direct way of obtaining global $\alpha$
which is different from the method discussed in \cite{pcm95}
for determining $\alpha_{best}$. The main differences are :
(1). the singularities at neutral line are automatically avoided
in our method by using the second moment of minimization and
(2). the computation of constant $\alpha$
force-free fields for different test values of $\alpha$
is not required.
\cite{hagi04} used a different parameter $\alpha_{av}^{(1)}$
to avoid the effect of Faraday rotation in sunspot umbrae. However,
modern inversion techniques using complete Stokes profiles are
free of this problem.

It must be noted that one can generate different values of
$\alpha_{g}$ using higher moments of minimization, e.g.,
by weighting $J_z$ with $B_z^n$, with n=3, 5, 7, ... etc.
The higher moments will be more sensitive to spatial
variation of $B_z$. Such large and complex variation
of $B_z$ is found generally in flare productive active
regions \citep{amba93,wang96,hagy99b}.
Thus we can try to use higher
order $\alpha_{g}$ as a global index for predicting the
flare productivity in active regions.

Finally, to compute $\alpha_{g}$ we need all the three components
of magnetic field which is obtained from the measurements of
vector magnetograms.
However, here we use the analytically generated bipole,
as discussed in the following section, with known values of all the
magnetic parameters to investigate the effect of polarimetric noise.

\section {Generation of theoretical bipole}

We use the analytic, non-potential force-free fields of the form
derived by \cite{low82}. These fields describe an isolated bipolar
magnetic region which is obtained by introducing
currents into a potential field structure.
This potential field is produced by an infinite straight line
current running along the
intersection of the planes y = 0 and z = -a, where negative sign
denotes planes below
the photosphere z = 0. At the photosphere (z = 0), the field has the
following form :
\begin{equation}\label{}
{B_x} = -\frac{B_0 a}{r}\cos\phi(r)
\end{equation}
\begin{equation}\label{}
{B_y} = \frac{B_{0}a x y}{{r} (y^2 + a^2)}\cos\phi({r})- \frac{B_0 a^2}{(y^2 + a^2)}\sin\phi(r)
\end{equation}
\begin{equation}\label{}
{B_z} = \frac{B_{0}a^2 x }{{r} (y^2 + a^2)}\cos\phi({r})- \frac{B_0 a y}{(y^2 + a^2)}\sin\phi(r)
\end{equation}

where $B_{0}$ is the magnitude of the field at origin and $ r^{2} = x^{2} + y^{2} + a^{2}$.
The function $\phi(r)$ is a free generating function related to the force-free
parameter $\alpha$ (see eqn (2)) by
\begin{equation}\label{}
    \alpha = -\frac{d\phi}{dr}
\end{equation}
which determines the current structure and hence the amount and location
of shear present in the region.
By choosing $\phi(r) = constant =\pi/2$ we can obtain a simple potential
(current-free, $\alpha = 0$) field produced by the infinite line
current lying outside the domain. Steeper gradient of $\phi(r)$ results
in a more sheared (non-potential) field.

In equation (11) the sign on the right hand side is taken positive in
 the paper by Low (1982) which is a typing mistake (confirmed by B. C.
 Low, private communication).
We mention this here to avoid carrying forward of this typo as was done in
\cite{wilk89}.

A grid of 100 x 100 pixels was selected for calculating the field
components.
The magnitude of field strength at origin has been taken as 1000G and
the value of `a' is
taken as 15 pixels (below the photosphere, z = 0).

The simulated field components with corresponding contours are shown
in the figure 1.

\begin{figure}
\centerline{\includegraphics[width=1.0\textwidth,clip=, ]{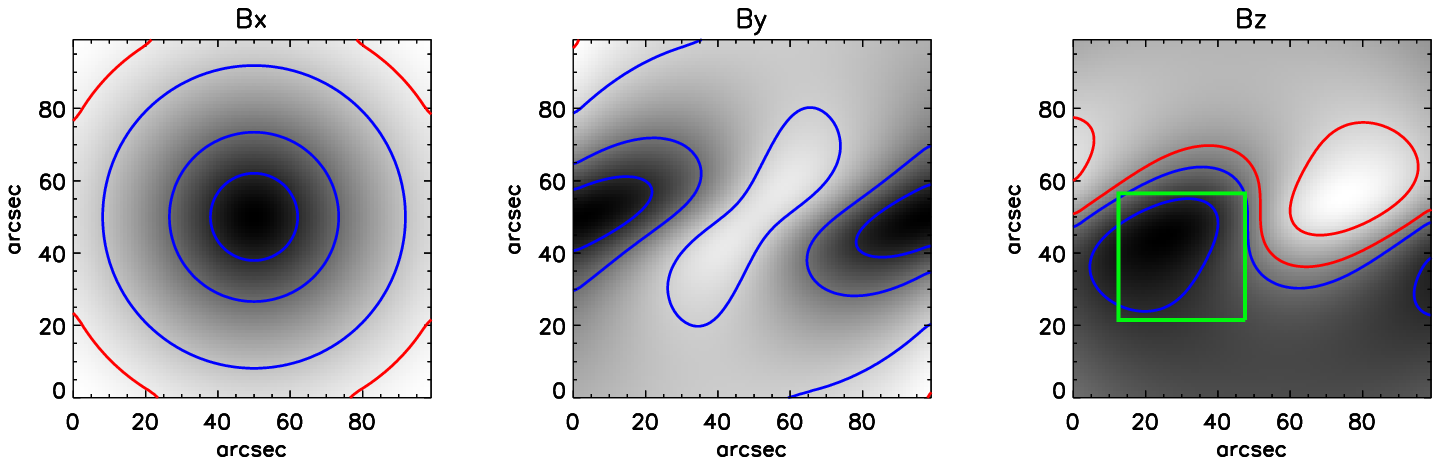}}
\caption{Contours of the field components overlaid on their gray-scale images.
The contour levels are  100,  500 and  800 G of magnetic fields. The red and blue
contours denote the `positive' and `negative' polarities, respectively.
The green box in B$_{z}$ shows the area which
is selected for the calculation of global $\alpha$.
For details see the text. \label{fig1}}
\end{figure}

Here we use the following function (e.g., \cite{wilk89})
for the generation of the field components ({B$_x$, B$_y$, B$_z$}) :
\begin{eqnarray}
  \phi(r) &=& \frac{\pi}{2}\frac{ r-a}{2 a}~,  \ \   r \leq 3 a \\
   &=& \frac{\pi}{2}~~~~~~~~, \ \ r > 3 a
\end{eqnarray}
Results for the fields generated by different $\phi(r)$ are
quantitatively similar.
In this way we generate a set of vector fields with known values of $\alpha$.

Most of the time one of the bipoles of a sunspot observed on the
Sun is compact (leading) and the other (following) is comparatively diffuse.
Observations of compact pole gives half of
the total flux of the sunspot and is mostly used for analysis.
One can derive the twist present in the sunspot using one
compact pole of the bipolar sunspot for constant $\alpha$.
Thus we have selected a single polarity of the analytical bipole
as shown in figure~\ref{fig1} to calculate the twist.

Fine structure in real sunspots is difficult to model.
Our analysis applies to the large scale patterns of the magnetic field
regardless of fine structure.

All the following sections discuss the analysis and results obtained.

\section{Profile generation from the analytical data and inversion}

Using the analytical bipole method \citep{low82} the non-potential
force-free field components B$_x$, B$_y$ \& B$_z$ in a plane have
been generated and are given as in equations (8), (9) \& (10).
We have shown B$_x$, B$_y$, \& B$_z$ maps (generated on a grid
of 100 x 100 pixels) in figure~\ref{fig1}.
From these components we have derived magnetic
field strength (B), inclination ($\gamma$) and azimuth
($\xi$ : free from 180$^o$ ambiguity).
In order to simulate the effect of typical polarimetric noise
in actual solar observations on magnetic field measurements
and study the error in the calculation
of $\alpha$ and magnetic energy, we have generated the synthetic
Stokes profiles for each B, $\gamma$ and $\xi$ in
a grid of 100 x 100 pixels, using the He-Line Information
 Extractor ``HELIX'' code \citep{lagg04}.
This code is a Stokes inversion code based on fitting the
observed Stokes profiles with synthetic ones obtained by
Unno-Rachkovsky solutions \citep{unno56,rach67}
to the polarized radiative transfer equations (RTE) under
the assumption of Milne-Eddington (ME) atmosphere
\citep{lando82} and local thermodynamical
equilibrium (LTE). However, one can also use this code for
generating synthetic Stokes profiles for an input model atmosphere.
The synthetic profiles are functions of magnetic field
strength (B), inclination ($\gamma$),
azimuth ($\xi$), line of sight velocity ($v_{Los}$),
Doppler width ($v_{Dopp}$), damping constant ($\Gamma$),
ratio of the center to continuum
opacity ($\eta_0$), slope of the source function ($S_{grad}$)
and the source function ($S_0$) at $\tau$ = 0.
The filling factor is taken as unity.
In our profile synthesis only magnetic field
parameters B, $\gamma$, $\xi$
are varied while other model parameters are kept
same for all pixels.
The typical values of other thermodynamical parameters are
given in table~\ref{tbl-1}. We use the same
parameters for all pixels. Further,
all the physical parameters at each pixel are taken to be
constant in the line forming region. However, one must remember
that real solar observations have often Stokes V area
asymmetries \citep{sol89,khom05} as a result of vertical
magnetic and velocity field gradients present in the line
forming region. This has not been taken into account in
our simulations.

\begin{figure}
\centerline{\includegraphics[width=1.0\textwidth,clip=,]{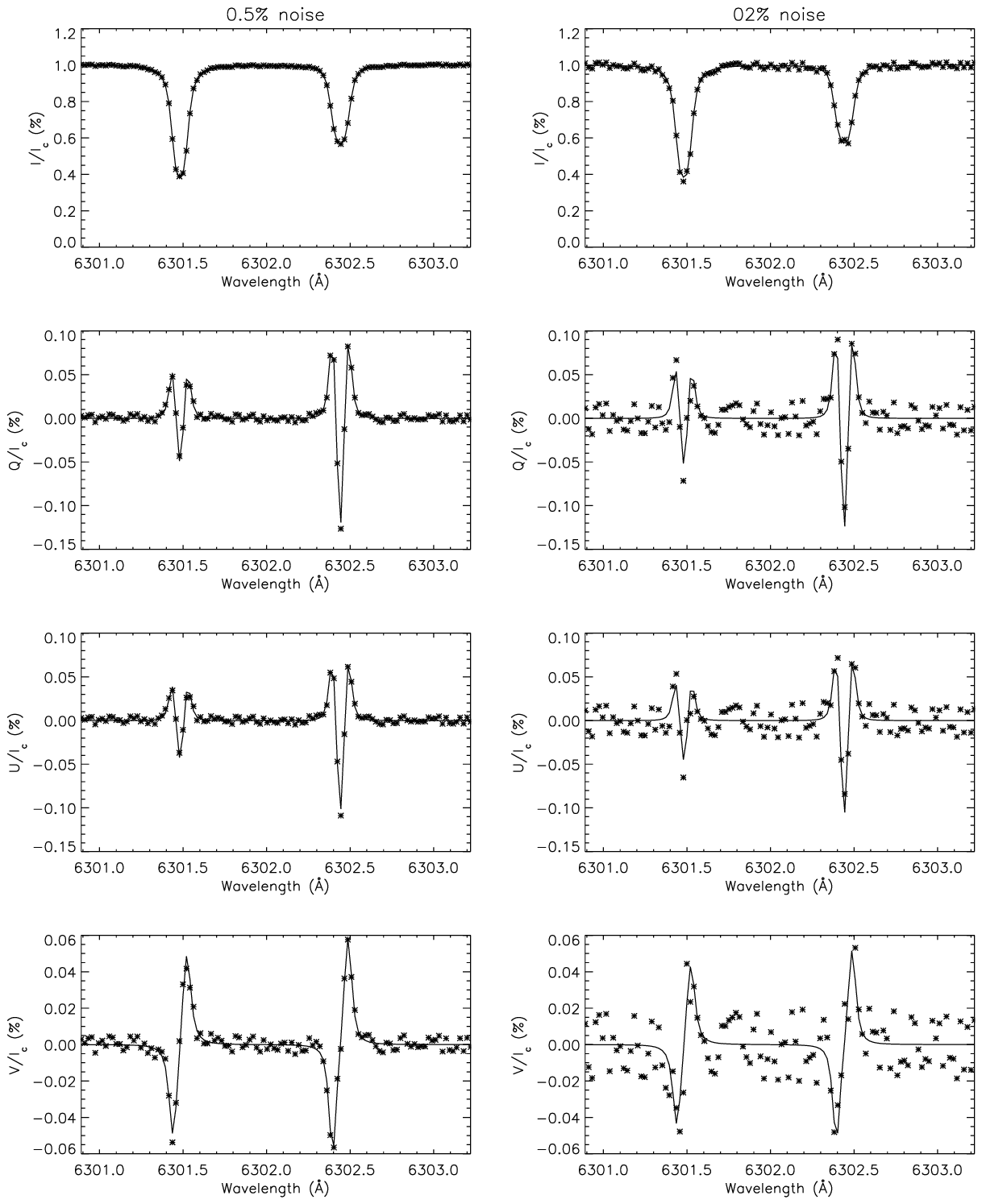}}
\caption{Example of Stokes profiles with 0.5\% (left column) and 2.0\%
(right column) noise along with fitted profiles.
The input parameters for the associated pixel are as follows : field strength=
861 G, inclination=101$^o$, azimuth=19$^o$. The corresponding output parameters
are 850 G, 101$^o$, 19$^o$ for 0.5\% noise and 874 G, 99$^o$, 19$^o$ for 2.0\%
noise. \label{fig2}}
\end{figure}

A set of Stokes profiles with 0.5\% and 2.0\% noise
for a pixel is shown in figure~\ref{fig2}.

\begin{figure}
   \centerline{\includegraphics[width=0.38\textwidth,clip=,]{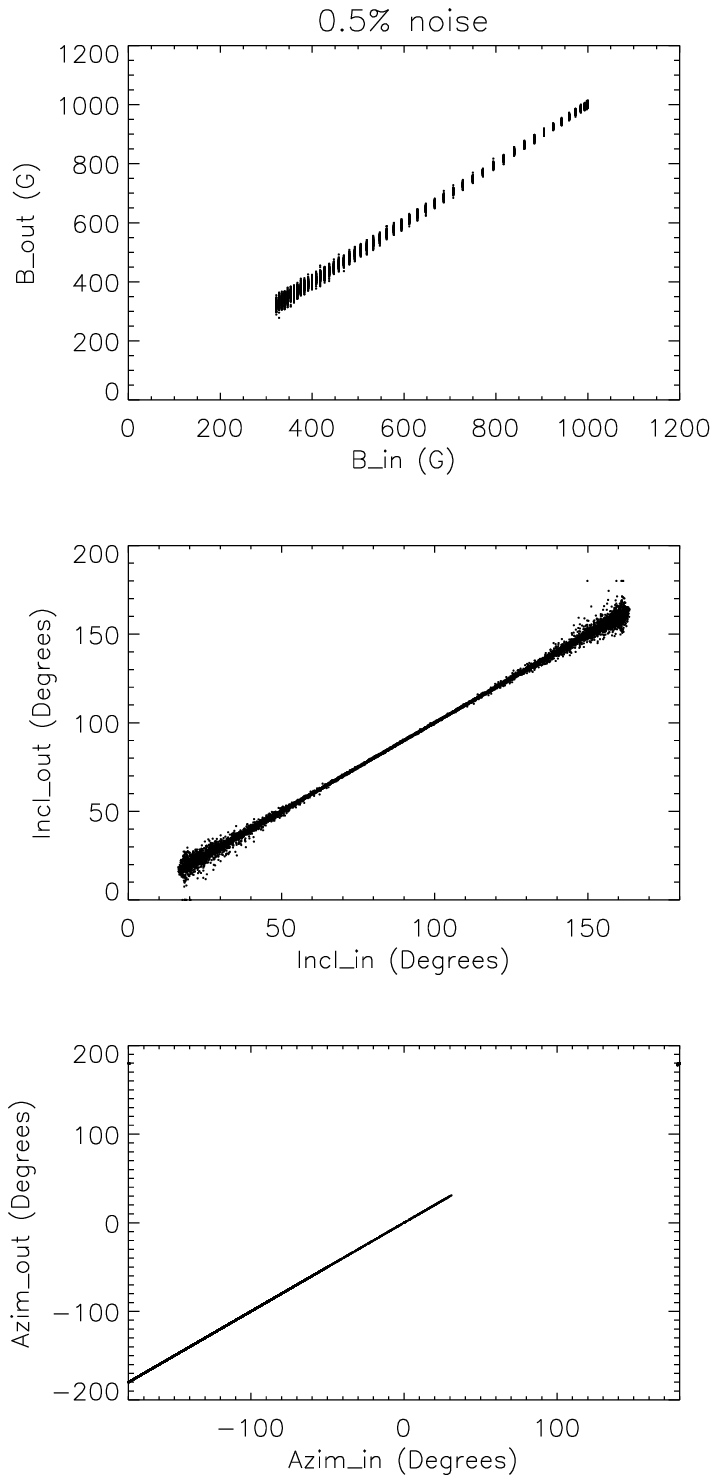}
           \includegraphics[width=0.38\textwidth,clip=,]{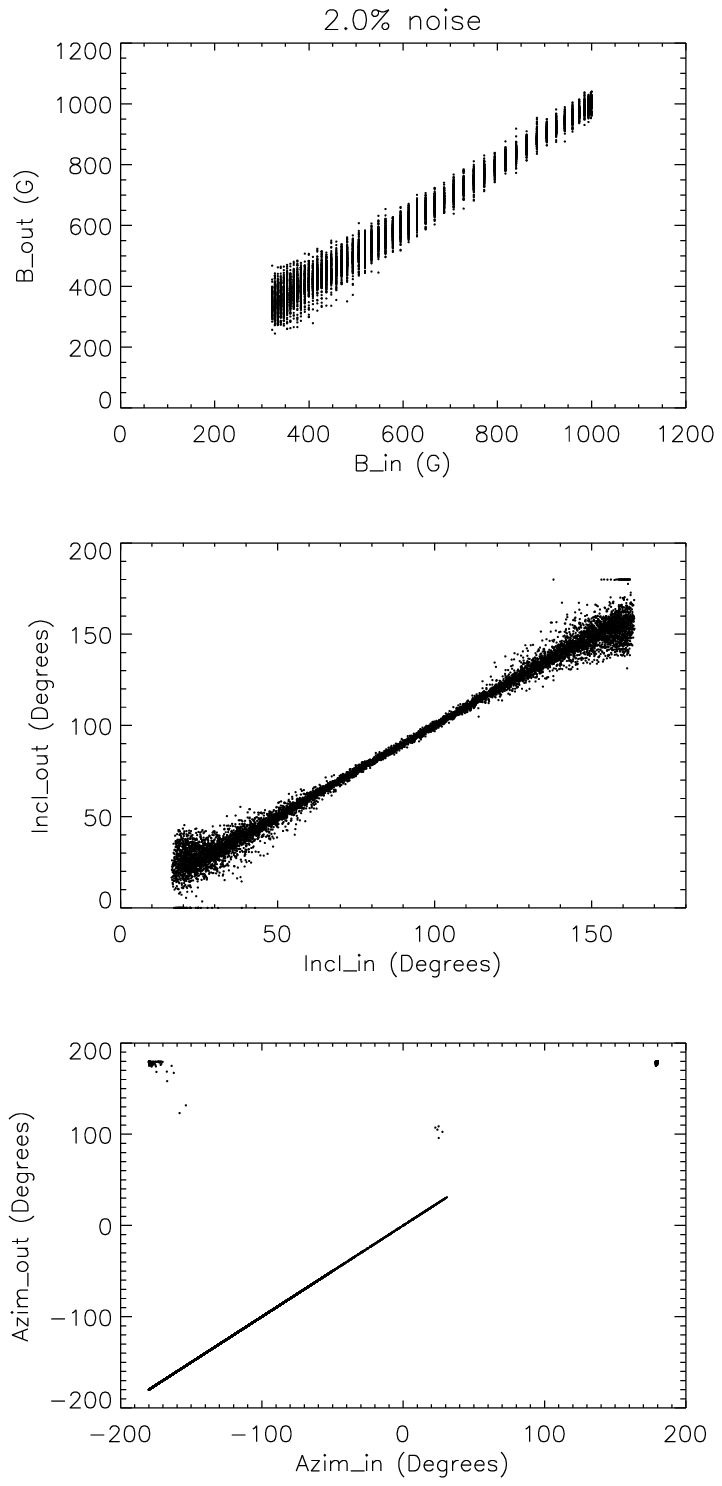}\\}
\centerline{\includegraphics[width=0.48\textwidth,clip=,]{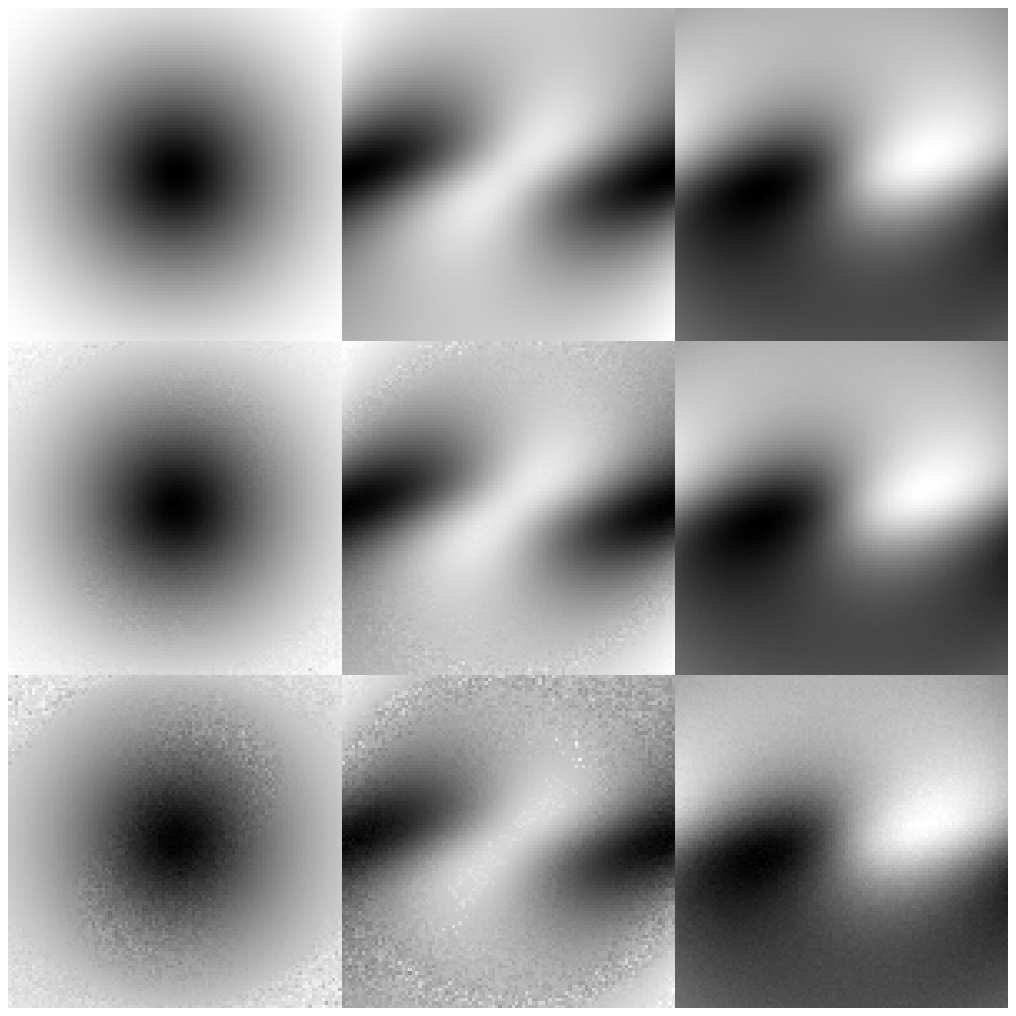}}
\caption{ Scatter plot (upper panel) between the field strength, inclination and azimuth
before and after inversion with 0.5\% (1$^{st}$ column) and 2.0\% (2$^{nd}$ column)
noises in the profiles.
The lower panel shows the images of vector fields { B$_{x}$, B$_{y}$ \& B$_{z}$} before
(1$^{st}$ row) and after inversion with 0.5\% (2$^{nd}$ row)
and 2\% (3$^{rd}$ row) noises in the profiles. \label{fig3}}
\end{figure}

The wavelength grid used for generating synthetic spectral profiles
is same as that of Hinode (SOT/SP) data which are as follows :
start wavelength of 6300.89 {\AA}, spectral sampling 21.5 m{\AA}/pixel,
and 112 spectral samples.
We add normally distributed random noise of different
levels in the synthetic Stokes profiles.
Typical noise levels in Stokes profiles obtained by Hinode (SOT/SP)
normal mode scan are of the order of 10$^{-3}$ of the continuum intensity,
I$_{c}$ \citep{ichi08}.
We add random noise of 0.5 \%
of the continuum intensity I$_{c}$ to the polarimetric profiles.
In addition, we also study the effect of adding a noise of
2.0\% level to Stokes profiles as a worst case scenario.
We add 100 realizations of the noise of the orders mentioned above
to each pixel and invert the corresponding 100 noisy profiles using the
``HELIX'' code.

The guess parameters to initialize the inversion are generated by
perturbing known values of B, $\gamma$
and $\xi$ by 10\%.
Thus after inverting 100 times we get 100 sets of B,
$\gamma$ \& $\xi$ maps for the input B,
$\gamma$ \& $\xi$ values from bipole data.
In this way we estimate
the spread in the derived field values for various field strengths,
inclinations etc. First, the inversion is done without adding any
noise in the profiles to check the
accuracy of inversion process. We get the results retrieved in
this process which are very similar to that of the initial
analytical ones. The scatter plot of input field strength,
inclination, azimuth against the corresponding retrieved
strength, inclination, azimuth after noise addition and
inversion is shown in figure~\ref{fig3} (upper panel).
Typical {B$_x$, B$_y$ \& B$_z$} maps with different noise
levels are shown in the lower panel.
As the noise increases {B$_x$, B$_y$ \& B$_z$} maps become
more grainy.

\begin{figure}
\centerline{\includegraphics[width=1.0\textwidth,clip=,]{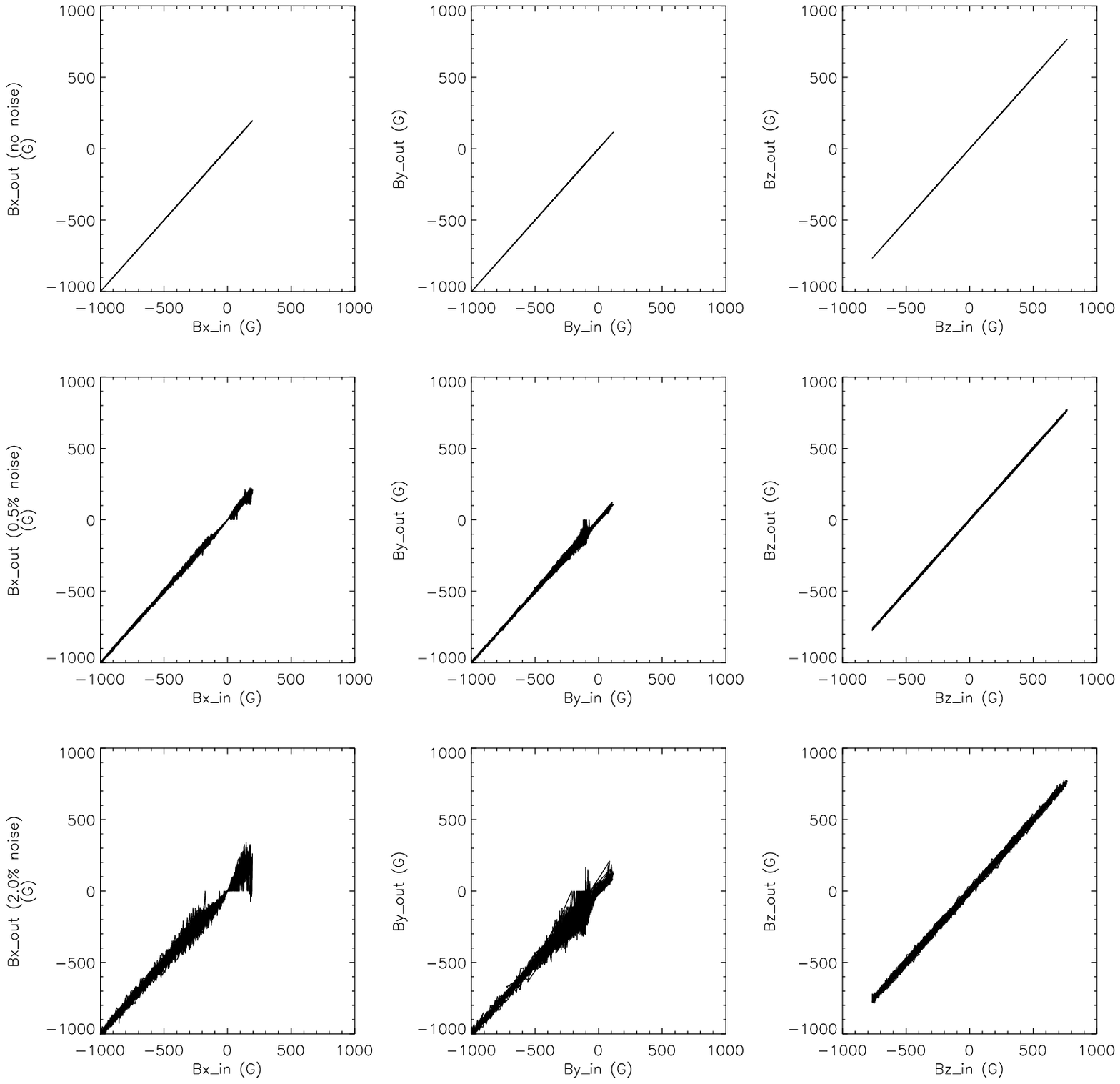}}
\caption{Scatter plot between { B$_x$, B$_y$ \& B$_z$} before and after
inversion without noise (1$^{st}$ row)
and with adding noise in the profiles: 2$^{nd}$ row with
0.5\% noise and 3$^{rd}$ row with 2.0\% noise (of I$_c$)
in the polarimetric profiles. \label{fig4}}
\end{figure}

From the plots shown in figure~\ref{fig3} we can see that the error in the
field strength for a given noise level decreases for strong fields.
This is similar to results of \cite{venk89}.
As the noise increases in the profiles, error in deriving the
field strength increases.
We find that the error in the field strength determination is $\sim 15\%$
for 0.5\% noise and $\sim 25\%$ for 2\% noise in the profiles.
Inclination shows more noise near 0$^o$ \& 180$^o$
than at $\thicksim 90^o$.
The error is less even for large noisy profiles
for the inclination angles between $\sim 50^o-130^o$.
The reason for this may be understood in the following way.
Linear polarization is weaker near 0 and 180$^o$ inclinations and is
therefore more affected by the noise.
The azimuth determination has inherent 180$^o$ ambiguity due to
insensitivity of Zeeman effect to orientation of transverse fields.
Thus in order to compare the input and output azimuths we resolve
this ambiguity in $\xi_{out}$ by comparing it with $\xi_{in}$ i.e.,
the value of $\xi_{out}$ which makes acute angle with $\xi_{in}$
has been taken as correct. We can see azimuth values after resolving
the ambiguity in this way show good correlation with
input azimuth values. Some scatter is due to the points where ambiguity
was not resolved due to 90$^o$ difference in $\xi_{in}$ and $\xi_{out}$.

First, the $\alpha_{g}$ was calculated from the vector field components
derived from the noise free profiles to
verify the method of calculating global alpha and also the inversion
 process. We have used the single polarity to calculate
global alpha present in sunspot as discussed in section 3.
We retrieved the same value of $\alpha_{g}$ as calculated
using the initial analytical field components.
From the figure~\ref{fig4}. we can see that the effect of noise
on the field components is not much
for the case of 0.5\% noise but as the noise in the profiles
 is increased to 2.0\%, the field components specially
 transverse fields show more uncertainty.
The vertical field is comparatively less affected with noise.
The scatter plot in figure~\ref{fig4}. shows that the inversion gives
good correlation to the actual field values.
The points with large scatter are due to poor ``signal to noise''
ratio in the simulated profiles.
The mean percentage error in the further discussions is given
in terms of weighted average of error.

\section{Estimation of the error in the calculation of global alpha ($\alpha_{g}$)}

We calculate the percentage error in global alpha each time after
getting the inverted results, for both
the cases when 0.5\% and 2.0\% (of I$_c$) noise is added in the profiles,
by the following relation :

\begin{equation}\label{}
\frac{\triangle \alpha_{g}}{\alpha_{g}}(\%) = \frac{\alpha_{g}^{*}-\alpha_{g}}
{\alpha_{g}}\times 100
\end{equation}
where $\alpha_{g}^*$ is calculated global alpha and
$\alpha_{g}$ is the analytical global alpha.

\begin{figure}    
  \centerline{\includegraphics[width=1.0\textwidth,clip=]{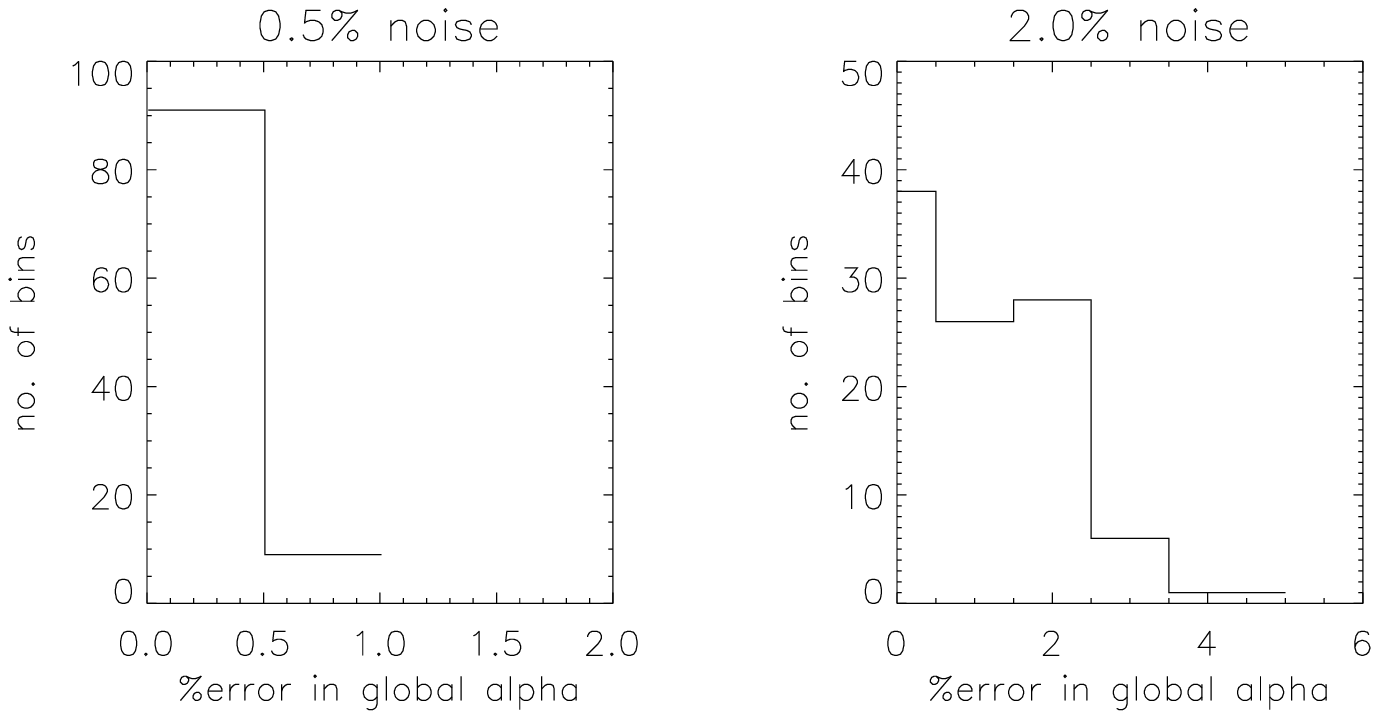}}
  \caption{Histogram of the percentage error in calculation of $\alpha_{g}$
  with 0.5\% and 2.0\% noise (of I$_{c}$) in polarimetric profiles,
  respectively. \label{fig5}}
  \label{F-simple}
 \end{figure}

The histogram of the results obtained is shown in figure~\ref{fig5}.

First, we inverted the profiles without adding any noise and calculated $\alpha_{g}$
from retrieved results to compare it with the `true' $\alpha_{g}$
calculated from the analytically generated vector field components. We get
less than 0.002\% difference in the both $\alpha_{g}$ values.

For the case of 0.5\% noise in polarimetric profiles we get a mean error of 0.3\% in
the calculation of $\alpha_{g}$ and error is never more than 1\%. Thus the
calculation of $\alpha_{g}$ is almost free from the effect of noise in this case.
Hence, by using data from Hinode (SOT/SP), one can derive the accurate value of
twist present in a sunspot.

If 2.0\% noise is present in the polarization, then maximum $\thicksim5\%$
error is obtained. Weighted average shows only 1\% error.
Thus the estimation of alpha is not influenced very much even from the data
obtained with old and ground based magnetographs.
In any event it is unlikely that a realistic error will be
large enough to create a change in the sign of $\alpha_g$.

\section {Estimation of the error in the calculation of magnetic energy (E$_{m}$)}

The magnetic energy has been calculated using virial theorem. One form of the general
virial theorem \citep{chandra61} states that for a force-free magnetic field,
the magnetic energy contained in a volume V is given by a surface integral over the
 boundary surface S,
\begin{equation}\label{}
    \int\frac{1}{8\pi}{B^2} dV = \frac{1}{4\pi}\int[\frac{1}{2} B^2{\bf r} -
    ({\bf B} \cdot {\bf r}){\bf B}]\cdot {\bf \hat{n}}\ \ dS
\end{equation}
where $\bf{r}$ is the position vector relative to an arbitrary origin, and \^{n} is
the normal vector at surface.
Let us adopt Cartesian coordinates, taking as z=0 plane for photosphere.
This assumption is reasonable because the size of sunspots are very small compared
to the radius of the Sun. If we make the further reasonable assumption that the
magnetic field strength decreases with distance more rapidly than r$^{-3/2}$
whereas a point dipole field falls off as r$^{-3}$, then the equation (15)
can be simplified to \citep{molo74}
\begin{equation}\label{}
\int\frac{1}{8\pi} {B^2} dx dy dz = \frac{1}{4\pi}\int(x B_{x} + y B_{y}) B_{z} dx  dy
\end{equation}
where x and y are the horizontal spatial coordinates. B$_x$, B$_y$ \& B$_z$
are the vector magnetic field components. This equation (16) is
referred as the ``magnetic virial theorem''.

Thus magnetic energy of an active region can be calculated simply by
substituting the derived vector field components into the surface
integral of equation (16) \citep{low82,low85,low89}.
Magnetic field should be solenoidal and force-free as is
the case for our analytical field. So the energy integral is
independent of choice of the origin.

If all the above conditions are satisfied then the remaining
source of uncertainty in the magnetic energy
estimation is the errors in the vector field measurements themselves.
So, before the virial theorem can be meaningfully applied to the Sun,
it is necessary first to understand how the errors in the vector
field measurements produce errors in the calculated
magnetic energies.

Earlier, the efforts were made to estimate the errors
\citep{gary87,klim92} for magnetographs like Marshall
Space Flight Center (MSFC) magnetograph.
\cite{gary87} constructed a potential field from MSFC data and computed
its virial magnetic energy. Then, they modified the vector field components by
introducing random errors in B$_{x}$, B$_{y}$ and B$_{z}$ and
recomputed the energy.
They found the two energies differ by 11\%.
\cite{klim92} approached
the problem differently. They introduced errors in the polarization
measurements from which the field is derived instead of introducing errors to
magnetic fields directly. In this way they were able to approximate reality,
more closely and were able to include certain type of errors such as crosstalk
which were beyond the scope of the treatment by \cite{gary87}.
 They found that the energy uncertainties are likely to
exceed 20\% for the observations made with
the vector magnetographs present at that time (e.g. MSFC).

Here, our approach is very similar to that of \cite{klim92}
except that we consider full Stokes profile measurements to derive the magnetic
fields like in the most of the recent vector magnetographs e.g., Hinode (SOT/SP),
SVM-USO etc. as mentioned earlier.
We begin with an analytical field, determine polarization signal as explained
in earlier parts, introduce the random noise of certain known levels
(0.5\% \& 2.0\% of I$_c$) in the polarization profiles, infer an `observed'
magnetic field after doing the inversion of the noisy profiles,
compute an `observed' magnetic energy from the `observed'
field and then compare this energy with the energy of the `true' magnetic field.
The percentage error is calculated from the following expression:
\begin{equation}\label{}
    \frac{\triangle E_{m}}{E_{m}}(\%) = \frac{E_{m}^{*}-E_{m}}{E_{m}}\times 100
\end{equation}
where E$_{m}^{*}$ is `observed' energy and E$_{m}$ is `true' energy.
All the above processes have been described in detail in section 4.

\begin{figure}    
  \centerline{\includegraphics[width=1.0\textwidth,clip=]{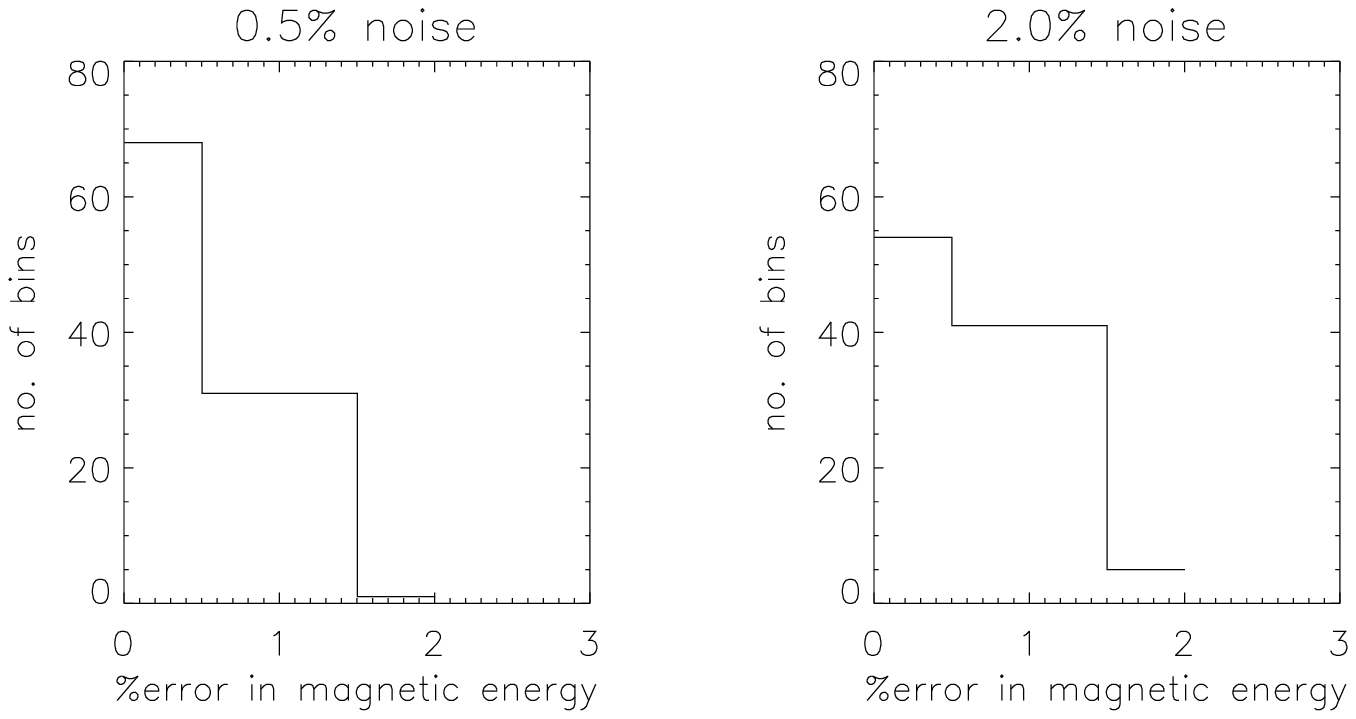}}
  \caption{Histogram of the percentage error in calculation of magnetic energy
  when 0.5\% and 2.0\% noise (of I$_{c}$) is present in polarimetric profiles,
   respectively. \label{fig6}}
\label{F-simple}
\end{figure}

Figure~\ref{fig6} shows the uncertainty estimated in the calculation
of magnetic energy in two cases when error in the polarimetric
profiles is 0.5\% and 2.0\% of I$_{c}$.
Needless to say, we first checked the procedure by calculating the
 magnetic energy from the vector fields derived
from inverted results with no noise in the profiles. We found the same
energy as calculated from the initial analytical fields.

We can see that the magnetic energy can be calculated with a very
good accuracy when less noise is present in the polarization as
is observed in the modern telescopes like Hinode (SOT/SP) for
which very small (of the order of 10$^{-3}$ of I$_c$) noise is
expected in profiles. We find that a mean of 0.5\% and maximum
up to 2\% error is possible in the
calculation of magnetic energy with such data.
So, the magnetic energy calculated from the Hinode data
will be very accurate provided the force-free field condition
is satisfied.

The error in the determination of
magnetic energy increases for larger levels of noise.
In the case of high noise in profiles (e.g.~2.0\% of I$_{c}$)
the energy estimation is very much vulnerable to the
inaccuracies of the field values. We replaced the inverted
value of the field parameters with the analytical value wherever
the inverted values deviated by more than 50\% of the `true'
values. We then get the result shown in the right panel. We can
see that the error is very small even in this case.
The mean value of error is $\sim0.7\%$.

\section{Discussion and Conclusions} 

We have discussed the direct method of estimating $\alpha_g$
from vector magnetograms using the $2^{nd}$ moment of minimization.
The higher order moments also hold promise for generating an index for
predicting the flare productivity in active regions.

The global value of twist of an active region can be measured with
a very good accuracy by calculating $\alpha_g$.
Accurate value of twist can be obtained even if one polarity of
a bipole is observed.

The magnetic energy calculation is very accurate as seen from our results.
Very less error (approximately 0.5\%) is seen in magnetic energy with
0.5\% noise in the profiles.
Thus we conclude that the magnetic energy can be
estimated with very good accuracy using the data obtained from modern
telescopes like Hinode (SOT/SP). This gives us the means to look for
magnetic energy changes released in weak C-class flares
which release radiant energy of the order of 10$^{30}$ ergs
(see Appendix-C), thereby improving the statistics.

These energy estimates are however subject to the condition that
the photospheric magnetic field is force-free, a condition which
is not always met with.
We must then obtain the energy estimates using vector magnetograms
observed at higher atmospheric layers where the magnetic field is
force-free \citep{metc95}.

The 180$^o$ azimuthal ambiguity (AA) is another source of error
for determining parameters like $\alpha_g$ and magnetic free energy
in real sunspot observations.
The smaller the polarimetric noise, the smaller is the uncertainty in
azimuth determination, thereby allowing us to extend the range of the
acute angle method used in our analysis. On the other hand it is
difficult to predict the level of uncertainty produced by AA.
Influence of AA is felt more at highly sheared regions which will
anyway deviate from the global alpha value. Thus, avoiding such pixels
will improve determination of $\alpha_g$.
Magnetic energy calculation at such pixels could be done by
comparing energy estimates obtained by `flipping' the azimuths
and choosing the mean of the smallest and the largest estimate
of the energy. Here we assume that half the number
of pixels have the true azimuth. This is the best one can do
for a problem that really has no theoretical solution allowed
by the Zeeman effect (but see also,
\cite{metc06} and references therein).
Observational techniques such as use of chromospheric chirality
\citep{lop06,mart08,tiwa08} or
use of magnetograms observed from different viewing angles
could perhaps resolve the AA .

Patches of both signs of alpha can be present in a
single sunspot \citep{pcm94,hagi04}.
In those cases the physical meaning of $\alpha_g$ becomes unclear.
Efforts are needed to understand the origin of such complex variation of
$\alpha$ in a sunspot.
Real sunspots show filamentary structures.
If this structure is accompanied by local variations of $\alpha$, then
does the global $\alpha$ result from correlations in the local $\alpha$
 values?
Or, are the small scale variations due to a turbulent cascade from the
large scale features?
The answers to these questions are beyond the scope of our present study.
Modeling sunspots with such complex fine structures is a great challenge.
However, we plan to address the question of fine structure of twists
in real sunspots observed from HINODE (SOT/SP), in our forthcoming study.

For the present, we demonstrate that the global twist present in an active
region can be accurately measured without ambiguity in its sign. Furthermore,
the high accuracy of magnetic energy estimation that can be obtained using
data from modern instruments will improve the probability for detecting
the flare related changes in the magnetic energy of active regions.

\acknowledgments
{\it \bf Acknowledgments }\\
\\
We thank Professor E. N. Parker for discussion leading to our understanding
about the physical meaning of $\alpha$ parameter during his visit to
Udaipur Solar Observatory in November 2007. We also thank him for looking
at an earlier draft of the manuscript and for making valuable
comments to improve it. One of us (Jayant Joshi) acknowledge financial
support under ISRO / CAWSES -  India programme. We thank Dr. A.
Lagg for providing the HELIX code. We are grateful for the
valuable suggestions and comments of the
referee which have significantly improved the manuscript.
\\

\appendix
 {\bf Appendix}
\section{Physical meaning of force-free parameter $\alpha$} 
(Derived from the discussions with Professor Eugene N. Parker during his
visit to Udaipur Solar Observatory)

Taking surface integral on both sides of eqn. (2), we get
\begin{eqnarray}
\nonumber  \alpha\int dS\cdot \bf B &=& \int dS \cdot \nabla \times \bf B \\
&=&  \oint dl \cdot {\bf B} ~~~~(from ~Stokes~ theorem)  \\
\nonumber or,\\
\alpha &=& \oint \frac{dl\cdot\bf B}{\Phi}
\end{eqnarray}


In the cylindrical coordinate we can write eqn.(A2) as
\begin{eqnarray}
 \nonumber \alpha &=& \frac{2\pi\varpi B_\Phi} {\pi \varpi^{2} B_{z}} \\
   &=& \frac{2 B_\phi}{\varpi  B_z}
\end{eqnarray}

where z and $\varpi$ are axial and radial distances from origin, respectively

The equation of field lines in cylindrical coordinates is given as :
\begin{equation}\label{}
    {\frac {B_z}{dz} = \frac{ B_\phi}{\varpi d \phi}}
\end{equation}
or,
\begin{equation}\label{}
    {\frac{ B_\phi}{ B_z} = \frac{\varpi d\phi}{dz}}
\end{equation}

Using eqns. (A3) \& (A5), we get
\begin{equation}\label{}
{\alpha = 2 ~\frac{ d \phi}{dz}}\\
\end{equation}

From equation (A6) it is clear that the $\alpha$  gives twice
the degree of twist per unit axial length.
If we take one complete rotation of flux tube i.e., $\phi= 2\pi$,
and loop length
\textsc{$\lambda \approx 10^{9} meters $}, then

\begin{equation}\label{}
    {\alpha = \frac{2\times 2\pi}{\lambda}}
\end{equation}
comes out of the order of  approximately $10^{-8}$ per meter.

\section{Correlation between sign of magnetic helicity and that of $\alpha$ }


Eqn. (2) can be written as
\begin{eqnarray}\label{}
\nonumber    \nabla \times \textbf{B} &=& \alpha (\nabla\times \bf A)\\
                             &=& \nabla\times (\alpha \bf A)
\end{eqnarray}
giving vector potential in terms of scalar potential $\phi$ as
\begin{equation}\label{}
    \bf A = \bf B \alpha^{-1} + \nabla \phi
\end{equation}
which is valid only for constant $\alpha$.\\
Using this relation in eqn.(1), we get magnetic helicity as
\begin{eqnarray}\label{}
\nonumber  H_{m} &=& \int ({\bf B} \alpha^{-1} + \nabla \phi) \cdot {\bf B} ~ dV \\
        &=& \int B^2 \alpha^{-1} dV + \int ({\bf B} \cdot \nabla) \phi ~ dV
\end{eqnarray}
Second term in the right hand side of eqn. (B3) can be written as,
\begin{eqnarray}\label{}
\nonumber \int ({\bf B} \cdot \nabla) \phi ~ dV &=& \int \nabla \cdot (\phi {\bf B}) ~ dV \\
&=& \int (\phi\ {\bf B})\cdot {\bf n}~ dS
\end{eqnarray}
(from Gauss Divergence Theorem)
which is equal to zero for a closed volume where magnetic field does not cross
the volume boundary ($\bf n \cdot B = 0$) provided that $\phi$ remains finite on the
surface.
Therefore, we get magnetic helicity in terms of $\alpha$ as
\begin{equation}\label{}
  H_{m} = \int B^2 \alpha^{-1}~ dV
\end{equation}
which shows that the force free parameter $\alpha$ has the same sign as
that of the magnetic helicity.
However, if $\bf n \cdot B \neq 0$, then the contribution of the
second term in eqn. (B3) remains unspecified.
Thus it is not correct to use alpha to determine the sign
of magnetic helicity for the half space above the
photosphere since $\bf n \cdot B \neq 0$ at the photosphere.

\section{Estimate of energy release in different classes of X-ray flares:}
With the simplifying assumption
that all classes of soft X-ray flares have a typical duration
of 16 min \citep{drak71}, we can see that the energy released
in the different classes of flares will be proportional
to their peak power. Since X-class flares typically
release radiant energy of the order of
$10^{32}$ ergs \citep{ems05}, therefore M-class, C-class, B-class
and A-class flares will release  radiant energy of the order of
respectively 10$^{31}$, 10$^{30}$, 10$^{29}$ and 10$^{28}$ ergs.

\label{S-appendix}

\begin{deluxetable}{lr}
\tabletypesize{\scriptsize}
\tablecaption{Model parameters for generating synthetic profile\label{tbl-1}}
\tablewidth{0pt}
\tablehead{
\colhead{Model Parameter} & \colhead{Value}}
\startdata
Doppler velocity, v$_{Los}$ (ms$^{-1}$)    &     0 \\
Doppler width, v$_{Dopp}$ (m{\AA})    &     20    \\
Ratio of center to continuum opacity, $\eta_o$      &     20         \\
Source function, S$_o$         &     0.001        \\
Slope of the source function, S$_{grad}$     &     1.0          \\
Damping constant, $\Gamma$      &     1.4          \\
\enddata
\end{deluxetable}


\end{document}